\documentstyle[epsf]{mn}
\newcommand{\x}{\mbox{$\times$}}
\newcommand{\e}[1]{\x10^{#1}}

\newcommand{\Mo}{\mbox{$M_\odot$}}

\newcommand{\Mbh}{\mbox{$M_{\rm bh}$}}

\newcommand{\Mw}{\mbox{$M_{\rm w}$}}
\newcommand{\Rw}{\mbox{$R_{\rm w}$}}

\newcommand{\dMw}{\mbox{$\dot{M}_{\rm w}$}}

\newcommand{\Rb}{\mbox{$R_\star$}}

\newcommand{\Nb}{\mbox{$N_\star$}}
\newcommand{\Mb}{\mbox{$M_\star$}}

\newcommand{\Lion}{\mbox{$L_{\rm ion}$}}

\newcommand{\rin}{\mbox{$r_{\rm in}$}}
\newcommand{\rout}{\mbox{$r_{\rm out}$}}
\newcommand{\fBS}{\mbox{$f_{\rm \/BS}$}}
\newcommand{\nBS}{\mbox{$n_{\rm \/BS}$}}

\def  \La          {\ifmmode {\rm Ly}\alpha \else Ly$\alpha$\fi}
\def  \Lalpha      {\ifmmode {\rm Ly}\alpha\,\lambda1215
		    \else Ly$\alpha$\,$\lambda1215$\fi}
\def  \Ka          {\ifmmode {\rm K}\alpha \else K$\alpha$\fi}
\def  \Lb          {\ifmmode {\rm L}\beta \else L$\beta$\fi}
\def  \Ha          {\ifmmode {\rm H}\alpha \else H$\alpha$\fi}
\def  \Halpha      {\ifmmode {\rm H}\alpha\,\lambda6563
		    \else H$\alpha$\,$\lambda6563$\fi}
\def  \Hb          {\ifmmode {\rm H}\beta \else H$\beta$\fi}
\def  \Hbeta       {\ifmmode {\rm H}\beta\,\lambda4861
		    \else H$\beta$\,$\lambda4861$\fi}
\def  \Pa          {\ifmmode {\rm P}\alpha \else P$\alpha$\fi}
\def  \HeI         {\ifmmode {\rm He}\,{\sc i}\,\lambda5876
		    \else He\,{\sc i}\,$\lambda5876$\fi}
\def  \HeII        {\ifmmode {\rm He}\,{\sc ii}\,\lambda1640
		    \else He\,{\sc ii}\,$\lambda1640$\fi}
\def  \CIII        {\ifmmode {\rm C}\,{\sc iii}\,\lambda977
		    \else C\,{\sc iii}\,$\lambda977$\fi}
\def  \CIIIb       {\ifmmode {\rm C}\,{\sc iii]}\,\lambda1909
		    \else C\,{\sc iii]}\,$\lambda1909$\fi}
\def  \CIV         {\ifmmode {\rm C}\,{\sc iv}\,\lambda1549
		    \else C\,{\sc iv}\,$\lambda1549$\fi}
\def  \bOIIIbA     {\ifmmode {\rm [O}\,{\sc iii]}\,\lambda4363
		    \else [O\,{\sc iii]}\,$\lambda4363$\fi}
\def  \bOIIIbB     {\ifmmode {\rm [O}\,{\sc iii]}\,\lambda5007
		    \else [O\,{\sc iii]}\,$\lambda5007$\fi}
\def  \OIIIb       {\ifmmode {\rm O}\,{\sc iii]}\,\lambda1663
		    \else O\,{\sc iii]}\,$\lambda1663$\fi}
\def  \OVb         {\ifmmode {\rm O}\,{\sc v]}\,\lambda1218
		    \else O\,{\sc v]}\,$\lambda1218$\fi}
\def  \OVI         {\ifmmode {\rm O}\,{\sc vi}\,\lambda1035
		    \else O\,{\sc vi}\,$\lambda1035$\fi}
\def  \OIVb        {\ifmmode {\rm O}\,{\sc iv]}\,\lambda1402
		    \else O\,{\sc iv]}\,$\lambda1402$\fi}
\def  \bOIIb       {\ifmmode {\rm [O}\,{\sc ii]}\,\lambda3727
		    \else [O\,{\sc ii]}\,$\lambda3727$\fi}
\def  \bOIb        {\ifmmode {\rm [O}\,{\sc i]}\,\lambda6300
		    \else [O\,{\sc i]}\,$\lambda6300$\fi}
\def  \NV          {\ifmmode {\rm N}\,{\sc v}\,\lambda1240
		    \else N\,{\sc v}\,$\lambda1240$\fi}
\def  \NIVb        {\ifmmode {\rm N}\,{\sc iv]}\,\lambda1486
		    \else N\,{\sc iv]}\,$\lambda1486$\fi}
\def  \NIIIb       {\ifmmode {\rm N}\,{\sc iii]}\,\lambda1750
		    \else N\,{\sc iii]}\,$\lambda1750$\fi}
\def  \MgII        {\ifmmode {\rm Mg}\,{\sc ii}\,\lambda2798
		    \else Mg\,{\sc ii}\,$\lambda2798$\fi}
\def  \bNeVb       {\ifmmode {\rm [Ne}\,{\sc v]}\,\lambda3426
		    \else [Ne\,{\sc v]}\,$\lambda3426$\fi}
\def  \NeVIII      {\ifmmode {\rm Ne}\,{\sc viii}\,\lambda774
		    \else Ne\,{\sc viii}\,$\lambda774$\fi}
\def  \SiIV        {\ifmmode {\rm Si}\,{\sc iv}\,\lambda1397
		    \else Si\,{\sc iv}\,$\lambda1397$\fi}
\def  \bFeXb       {\ifmmode {\rm [Fe}\,{\sc x]}\,\lambda6734
		    \else [Fe\,{\sc x]}\,$\lambda6734$\fi}
\def  \bFeXIb      {\ifmmode {\rm [Fe}\,{\sc xi]}\,\lambda7892
		    \else [Fe\,{\sc xi]}\,$\lambda7892$\fi}
\def  \FeII        {\ifmmode {\rm Fe}\,{\sc ii}\,
		    \else Fe\,{\sc ii}\,\fi}
\def  \MgIIw       {\ifmmode {\rm Mg}\,{\sc ii}\,\lambda2798{\rm+}\FeII
		    \else Mg\,{\sc ii}\,$\lambda2798$+$\FeII$\fi}
\def  \MgIIc       {\ifmmode {\rm Mg}\,{\sc ii}\,\lambda2798{\rm-}\FeII
		    \else Mg\,{\sc ii}\,$\lambda2798$-$\FeII$\fi}

\load{\normalsize}{\sc}

\begin{document}

\title[Bloated Stars as AGN Broad Line Clouds]
      {Bloated Stars as AGN Broad Line Clouds:\\
      The Emission Lines Response to Continuum Variations}
\author[T. Alexander]
      {Tal Alexander\thanks{E-mail address: tal@wise.tau.ac.il}\\
       School of Physics and Astronomy, 
       Tel-Aviv University, 
       Tel-Aviv 69978, Israel.}
\maketitle 
\begin{abstract} 
The `Bloated Stars Scenario' proposes that AGN broad line emission
originates in the winds or envelopes of bloated stars (BS). Alexander and
Netzer (1994, 1996) established that $\sim 5\e{4}$ BSs with dense,
decelerating winds can reproduce the observed emission line spectrum and
line profiles while avoiding rapid collisional destruction.  Here, I
investigate a third prediction of the model, related to the size of the
broad line region, by deriving the emission line response to variations in
the ionizing continuum (`line reverberation') and comparing it to
observations. The expected time lags, as well as the order of response of
the various lines, strongly depend on the typical variability time scale of
the ionizing continuum. The BS model studied here corresponds to a bright
Syfert 1 galaxy (Sy1) or a low luminosity QSO. I find that the BS model is
consistent with the observed correlation between the Balmer lines time lags
and the AGN luminosity, which at present is the only line reverberation
information available for this luminosity class. The model displays also
some of the anti-correlation between the time lags of the metal emission
lines and their degree of ionization that has been observed in a few
low-luminosity Sy1s. However, the bright Sy1 model results differs from the
low-luminosity Sy1 data in that the $\Hb$ time lag is relatively shorter
and the $\CIV$ time lag relatively longer.

\end{abstract}

%%%%%%%%%%%%%%%%%%%%%%%%%%%%%%%%%%%%%%%%%%%%%%%%%%%%%%%%%%%%%%%%%%%%%%%%
%%%%%%%%%%%%%%%%%%%%%%%%%%%%%%%%%%%%%%%%%%%%%%%%%%%%%%%%%%%%%%%%%%%%%%%%
\section{Introduction}
  
    The observed properties of active galactic nuclei (AGN) lead to the
conclusion that the broad line emission originates in numerous small, cold
and dense gas concentrations, which are photoionized by the central
continuum source. These objects are labeled ``clouds'', although their true
nature remains unknown. A basic challenge for any broad line region (BLR)
model is to specify the physical mechanism that protects the clouds from
rapid disintegration in the AGN's extreme environment, or else specify a
source for their continued replenishment. The bloated stars (BSs) model
\cite{Edwards,Mathews,SN,Penston,Kazanas} proposes that the lines are
emitted from the winds or mass loss envelopes of giant stars. The star
provides both the gravitational confinement and the mass reservoir for
replacing the gas that evaporates from the envelope to the interstellar
medium (ISM). The BS model is further motivated by the lack of
observational evidence for net radial motion in the BLR (e.g. Maoz et
al. 1991, Wanders et al. 1995). This is consistent with virialized motion
in the gravitational potential of the nucleus.

   The BS model was recently studied by Alexander \& Netzer \shortcite{AN1}
(paper I) and Alexander \& Netzer \shortcite{AN2} (paper II), which used a
detailed photoionization code to calculate the emission line spectrum and
profiles of various BS wind models. The integrated emission line spectrum
was calculated by combining the line emissivity of the BSs with theoretical
models of the stellar distribution function. The model results were
compared to the mean AGN line ratios and to estimates of the BLR size and
line reddening. The mean observed $\La$ equivalent width was used to
determine the number of BSs and their fraction in the stellar population
and to estimate the collisional mass loss rate and its effect on the ISM
electron scattering optical depth. Detailed models of the stellar velocity
field were then used to derive the line profiles, which in turn constrained
the BS distribution within the stellar population.

   The main result of papers I and II is that there are BS wind models that
can reproduce the emission line ratios and profiles to a fair degree with
only a small fraction of BSs in the stellar population. However, these
models do not resemble winds of normal supergiants. The photoionization
calculations show that the emission line spectrum is dominated by the
conditions at the outer boundary of the line emitting zone of the wind. The
successful BS models are those with dense envelopes that have small density
gradients. In this case, the wind boundary is set by tidal forces near the
black hole and by the finite mass of the wind at larger radii. Only $\sim
5\x10^4$ such BSs ($< 1\%$ of the BLR stellar population) are required for
reproducing the BLR emission. As a result, the collision rate is reasonably
small ($< 1 \Mo$/yr). On the other hand, lower-density wind models or those
with steep density gradients are ruled out because they emit strong broad
forbidden lines, which are not observed in AGN. In addition, their low line
emissivity requires orders of magnitude more BSs in the BLR. This leads to
rapid collisional destruction, very high mass loss rate and very short BS
lifetimes.  Even if the BSs are continuously created, the ISM electron
scattering optical depth is very large, contrary to what is observed.

   Paper II shows that in order to reproduce the typical shapes and widths
of the observed emission line profiles it is necessary to assume that the
BS fraction in the stellar population falls off as $r^{-2}$. The emerging
picture is that AGN harbour an exotic class of giant stars, whose existence
is linked to the uniquely violent environment of AGN, and are consequently
much rarer in the more usual stellar environments outside the inner
nucleus. There are at least two astronomical precedences for such a situation:
The existence of blue stragglers in the high density cores of stellar
clusters, which is thought to be linked to stellar mergers or collisions
(e.g. Bailyn 1995) and the recently discovered concentration of extremely
rare He\,{\sc i} blue giants in the inner 1 pc of the Galactic Center
\cite{KGDR}. The envelopes of these blue giants reach densities as high as
those assumed in the BS model \cite{NKKGLH}.

   The present work explores a third aspect of the BS model: BLR line
reverberation, or echo mapping.  The continuum luminosity of AGN is
observed to vary in time and in many cases, the continuum variability
pattern is echoed after some time lag by the line fluxes. This is
considered the most direct support for the photoionization hypothesis of
BLR emission. The time lag between the continuum and line variations is
interpreted as the difference between the light travel time along the
direct line of sight and that along the indirect path from the continuum
source, via the line emitting gas, to the observer. The continuum and line
light curves are therefore related by the spatial emissivity distribution
(`emissivity map') of the line, which in turn depends on the distribution
of line emitting objects and their ionization structure.

   Extracting information on the BLR geometry from line reverberation data
is severely limited by observational difficulties and the ensuing
ambiguities in the statistical analysis (e.g. Maoz 1994). A reliable
analysis requires good sampling over a long period, a condition which is
seldom met. Only a few monitoring campaigns were carried out to date, each
with a different duration and sampling frequency. In contrast to the case
of the line ratios and profiles, very little can be determined at this
stage about the statistics of the time lags and their properties. However,
some general trends are beginning to emerge: the BLR size seems to be
positively correlated with the continuum luminosity, and high ionization
metal lines seem to originate at smaller radii than the low ionization ones
\cite{Clavel}.

   In this third part of the work I study the line reverberation properties
of a BS model with a luminosity of a bright Seyfert 1 galaxy (Sy1) or a
low-luminosity QSO. This model succeeded in reproducing the line ratios and
profiles of a typical AGN, and here I check whether it is also consistent
with the limited available line reverberation data. Section \ref{s:model}
briefly summarises the main components of the BS model. Section
\ref{s:calc} describes the line reverberation simulation procedure. The
simulated results are presented in section \ref{s:result} and discussed in
section~\ref{s:discuss}.

%%%%%%%%%%%%%%%%%%%%%%%%%%%%%%%%%%%%%%%%%%%%%%%%%%%%%%%%%%%%%%%%%%%%%%%%
%%%%%%%%%%%%%%%%%%%%%%%%%%%%%%%%%%%%%%%%%%%%%%%%%%%%%%%%%%%%%%%%%%%%%%%%

\section{The BS model}
\label{s:model}

    As described in detail in papers I and II, t he BS model is specified by
two main ingredients. The stellar distribution around the black hole and
the density structure of the BS envelope. The BS model that is discussed
in this paper is model C of paper II.

   The stellar distribution function that I use is based on the numeric
results of Murphy, Cohn \& Durisen \shortcite{MCD} (MCD), which follow the
evolution of a spherically symmetric multi-mass coeval stellar cluster in
the presence of a central black hole. The black hole mass grows as it
accretes mass from the stars, both directly, by tidal disruption, and
indirectly, from mass loss in the course of stellar evolution and of
inelastic stellar collisions. While the calculations deal in a
self-consistent way with the stellar dynamics and the build-up rate of the
central mass, the resulting AGN luminosity is not well defined. This is
because the calculations of the nucleus evolution are insensitive to the
properties of the central mass, such as the ratio between the black hole
mass and the infalling mass or to the black hole's angular momentum, nor
are they affected by the details of the accretion mechanism or by the
rest-mass to luminosity conversion efficiency, $\eta$. MCD make the simple
assumption that the bolometric luminosity is $L_{\rm bol}(t) = \eta
\dot{M}(t)c^2$ with $\eta = 0.1$, where $\dot{M}$ is the total stellar mass
loss rate from the stellar system. In this work I use MCD model 2B, at $t_0
= 10^9$ yr, when the black hole mass is $\Mbh = 1.9\x10^8\Mo$, and the
ionizing luminosity $L_{\rm ion} = 3.6\x10^{44}$ erg s$^{-1}$. This
luminosity is sub-Eddington and is determined by the total mass loss rate
from the stellar system. This evolved system represents a luminous Sy1 or a
low-luminosity QSO.  Once the choice of the stellar model is made, the
remaining free parameters are $\fBS$, the fraction of BSs in the normal
stellar population and the inner and outer radii of the BLR, $\rin$ and
$\rout$. In paper II, it was shown that these parameters are constrained by
the observed emission line profiles to values of roughly $\rin = 0.001$ pc,
$\rout = 0.25$ pc, and $\fBS \propto r^{-2}$, with the BSs comprising less
than 1\% of all stars.

   A single BS is modeled by a simplified, spherically symmetric structure
with no claim to hydrodynamical self-consistency and without specifying the
process that drives the wind. It has two components: a giant star of radius
$\Rb = 10^{13}$ cm and a mass of $\Mb = 0.8\Mo$, emitting a spherically
symmetric wind at a rate of $\dMw = 10^{-6} \Mo$/yr. The wind, or envelope,
extends to a radius $\Rw$ which contains an additional mass of up to $\Mw =
0.2\Mo$. The wind density is parameterized as a power-law, with two free
parameters
\begin{equation} 
N(R) = \Nb(R/\Rb)^{-\beta}\,,
\end{equation}
where $\Nb$ is the hydrogen number density at the base of the envelope. 
The combined constraints of the observed line ratios and profiles fix these
parameters to values of roughly $\beta = 3/2$ and $\Nb = 3.8\e{11}$
cm$^{-3}$. The proximity of the BSs to the black hole makes tidal
disruption the main physical mechanism that limits the size of the wind.
For example, at 30 ld, the BS wind radius (model C) is $\Rw \sim 2\e{14}$
cm.

   The BS model is fully specified by the choice of the AGN dynamical age,
$t_0$, which fixes $\Mbh$, by the BS structure and the ionizing continuum
spectrum. However, tidally-limited BSs obey certain approximate scaling
relations that link $\Nb$ and $\Lion$ in a way that allows $\Lion$ and
$\Nb$ to be jointly modified without affecting the emissivity map (paper
II). Since the emissivity map determines the line ratios and profiles, this
scaling property means that the model luminosity cannot be currently fixed
by either the BS model or the MCD AGN model. This introduces some
uncertainty in matching the BS model with observed AGN that complicates the
comparison of the model time lags with the observed
ones~(section~\ref{s:discuss}).

%%%%%%%%%%%%%%%%%%%%%%%%%%%%%%%%%%%%%%%%%%%%%%%%%%%%%%%%%%%%%%%%%%%%%%%%
%%%%%%%%%%%%%%%%%%%%%%%%%%%%%%%%%%%%%%%%%%%%%%%%%%%%%%%%%%%%%%%%%%%%%%%%

\section{Calculations}
\label{s:calc}

   In order to calculate the light curves of the various emission lines it
is necessary to specify the line response to the local changes in the
ionizing flux. A commonly assumed simplification is that the response is
linear, in which case the relation between the line light curve, $L(t)$,
and the ionizing continuum light curve $C(t)$ can be described by a
convolution with a transfer function $\Psi$ \cite{BM}
\begin{equation}
L(t) = \int_{-\infty}^{+\infty}C(\tau)\Psi(t-\tau)d\tau\,.
\label{e:Lt}
\end{equation}
In the case of a spherically symmetric BLR the transfer function is
\cite{Maoz2}
\begin{equation} 
\Psi(t) \propto \left\{ 
\begin{array}{lrcl}
	                    0, &            & t & < 0      \\[5pt]
\int_{\rin}^{\rout} w(r)c/2r dr,  &      0 \le & t & < 2\rin/c \\[5pt]
\int_{ct/2}^{\rout} w(r)c/2r dr, & 2\rin/c \le & t & < 2\rout/c \\[5pt]
                            0, & 2\rout/c \le & t &
\end{array}
          \right.
\label{e:psi}
\end{equation} 
The weight $w(r)$ is the normalized line emissivity from a shell
of BSs at radius $r$,
\begin{equation} 
w(r) = r^2\nBS(r)L_\ell(r)\left/ 
       \int_{\rin}^{\rout}r^2\nBS(r)L_\ell(r) dr\,,\right.
\end{equation} 
where $L_\ell(r)$ is the BS emissivity in line $\ell$ at a distance $r$
from the continuum source and $\nBS$ is the BS spatial density. $w(r)$ is
calculated by the photoionization code for the specific BS model.  The
assumption of a linear response implicitly includes the approximation that
the varying continuum does not change the emissivity map of the BLR,
i.e. that $\Psi$ is unaffected by $C(t)$. While this is not strictly true
(the ionization parameter is proportional to the ionizing flux), it is
probably an adequate approximation as long as the variations are not large.
Another way of interpreting $\Psi(t)$ is that it is the line response to a
$\delta$-function continuum flash.

\begin{figure} 
   \centering \epsfxsize=240pt \epsfbox{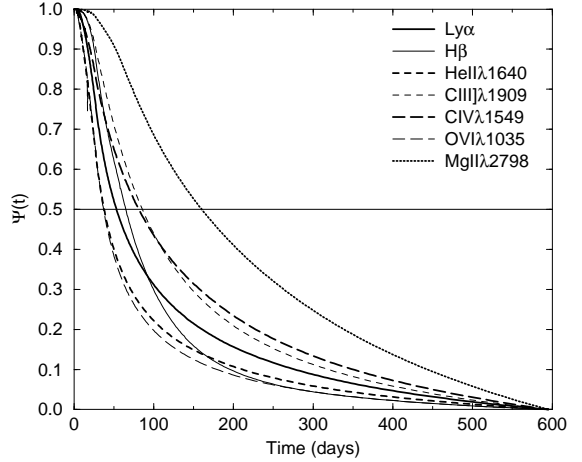}
   \caption{The transfer functions $\Psi(t)$ of various lines as calculated
   for model C. $\Psi$ is normalized to 1 at $t=0$.}
\label{f:psi}
\end{figure}

   As is the standard practice in the field of line reverberation studies,
I identify the time lag between the two light curves with the peak in
their cross-correlation function (CCF) \cite{BM}.

%%%%%%%%%%%%%%%%%%%%%%%%%%%%%%%%%%%%%%%%%%%%%%%%%%%%%%%%%%%%%%%%%%%%%%%%
%%%%%%%%%%%%%%%%%%%%%%%%%%%%%%%%%%%%%%%%%%%%%%%%%%%%%%%%%%%%%%%%%%%%%%%%

\section{Results}
\label{s:result}

   The present calculations demonstrate the general behaviour of the BS
system by describing in detail a particular model that was investigated in
paper II. This model (model C) is characterized by a relatively low
$\Lion/\Mbh$ ratio ($3.6\x10^{44}$ erg s$^{-1}$ / $1.9\x10^8\Mo$), a BS
fraction in the stellar population that decreases like $r^{-2}$ and a
relatively low hydrogen density at the base of the wind ($3.8\e{11}$
cm$^{-3}$). I calculated the transfer functions of various emission lines,
which are displayed in Fig.~\ref{f:psi}, by inserting the photoionization
results for model C into equations~\ref{e:Lt} and \ref{e:psi}.

   There is a distinct difference between the transfer functions of the
high ionization lines $\HeII$ and $\OVI$, and that of the low ionization
line $\MgII$. This reflects the fact that in this model, the high ionization
lines are formed at the inner BSs, where the ionizing flux is high, whereas
the low ionization lines are formed further away from the continuum source,
where the lowered ionizing flux allows the existence of extended, partially
ionized regions in the BS envelope. This trend resembles what is found in
line reverberation campaigns, namely that the high ionization metal lines
have shorter time lags than the low ionization metal lines.

\begin{figure} 
   \centering \epsfxsize=240pt
   \epsfbox{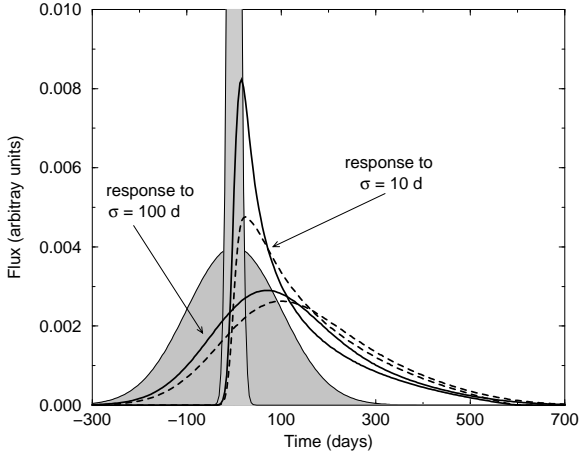} 
   \caption{The response of the model C $\La$ (bold line) and $\MgII$
   (dashed bold line) lines to narrow ($\sigma = 10$ d) and wide ($\sigma =
   100$ d) Gaussian flashes (shaded areas), which are centered on t = 0.}
   \label{f:gwidth}
\end{figure}

   The time lag of a line increases with the width of its transfer
function. However, the relative widths of the transfer function give only a
rough indication of the rank order of the line lags. The time lag of an
extended BLR geometry, such as that of model C, depends also very strongly
on the time-scale of the ionizing continuum, because the line CCF is the
convolution of $\Psi$ with the continuum auto-correlation function
\cite{Penston2,Netzer}. I check this here by calculating the response of the
lines to Gaussian shaped continuum flashes of various durations
(parameterized by their standard deviation $\sigma$). Figure~\ref{f:gwidth}
shows two such flashes with duration of $\sigma = 10$ and 100~d,
respectively, and the responses of the $\La$ and $\MgII$ lines. As
expected, the narrow, $\sigma = 10$~d flash induces a line light curve very
similar in shape to its transfer functions and consequently, the two light
curves respond quickly at lags of 20~d and 32~d, respectively. On the other
hand, the wide, $\sigma = 100$~d flash results in noticeably larger time
lags of 84~d and 120~d, respectively, and a larger lag difference between
the two lines.

\begin{figure} 
   \centering 
   \epsfxsize=240pt
   \epsfbox{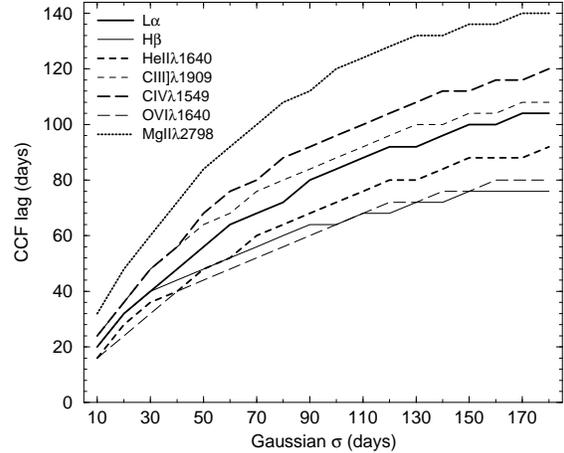} 
   \caption{The response of various model C lines to Gaussian shaped 
   continuum flashes, as a function of the Gaussian standard deviation 
   $\sigma$.}
   \label{f:gcomp}
\end{figure}

   The response times of various lines as a function of the Gaussian pulse
width are summarized in Fig.~\ref{f:gcomp}. Note that even for pulses as
wide as $\sim 1$ year, most of the line lags continue to increase with the
continuum time-scale. Only the responses of $\Hb$ and $\OVI$ appear to
begin to saturate. The different rates of time lag increase result in the
surprising effect that not only the time-lags, but also their rank order,
changes with increasing continuum variability time scale. For example, the
$\Hb$ line, which lags after the $\OVI$ and $\HeII$ lines when the ionizing
continuum varies quickly, precedes these lines when the continuum varies
slowly.  In some lines the time lag is a steep function of the continuum
variability timescale. For example, the $\HeII$ line time lag increases by
a factor of 6 when the width of the continuum flash is increased by a
factor of 18.  Although I demonstrated these trends only in the case of
one particular BS model and single Gaussian flashes, I expect these
results to hold also in the general case. This is because all BS models
have extended BLR geometries and all realistic continuum light curves can
be approximated, to some degree, by a superposition of Gaussian flashes.

\begin{figure} 
   \centering 
   \epsfxsize=240pt
   \epsfbox{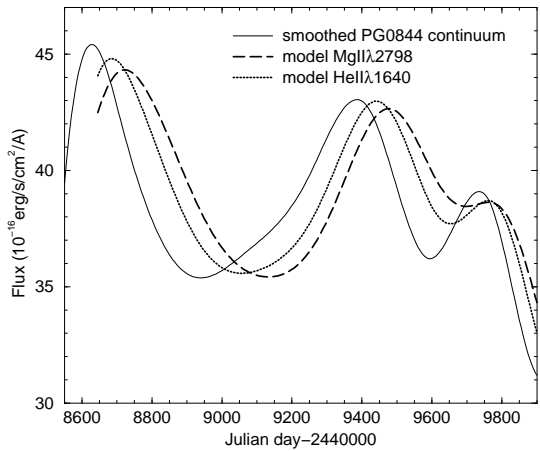} 
   \caption{The response of model C $\HeII$ and $\MgII$ lines to the
   smoothed 4759\AA\/ (rest frame) continuum of PG0844 (Kaspi 1996, private
   communication). In order to make the comparison between the light curves
   easier, the displayed line light curves are shifted to the continuum
   mean and their variation is scaled up to the continuum r.m.s variation.}
   \label{f:PG}
\end{figure}

   A realistic ionizing continuum is made up of many peaks of different
widths. I model this by using the recently obtained long light curve of
PG0844 (Kaspi 1996, private communication). Figure~\ref{f:PG} shows the
smoothed observed PG0844 optical continuum and the response of the $\HeII$
and $\MgII$ lines. The lags were calculated without the first 100 days of
the line light curves, which is roughly the time it takes $\Psi$ to fall to
half its initial maximal value. At times shorter than that, the line light
curve depends strongly on the preceding, unobserved continuum and cannot be
calculated reliably.  Note the difference between the lines response to the
wide second continuum peak and their response to the narrow third
peak. This is the pattern of behaviour that was suggested by the
simulations with single Gaussian flashes and by the simulations of P\'erez,
Robinson \& de la Fuente \shortcite{PRF}. Such a behaviour was actually
observed in the 1989 NGC 5548 IUE campaign \cite{NM}. Table~\ref{t:lags}
lists the lags of all the lines calculated here. As with the Gaussian
flashes, there is an anti-correlation between the time lag and the degree
of ionization.

\begin{table} 
\centering
\begin{tabular}{lcc}
\hline
Line      &lag (days) &rank order\\
\hline
$\La$     & 70        & 3        \\
$\Ha$     & 75        & 4        \\
$\Hb$     & 60        & 1        \\
$\HeI$    & 70        & 3        \\
$\HeII$   & 65        & 2        \\
$\CIII$   & 75        & 4        \\
$\CIIIb$  & 80        & 5        \\  
$\CIV$    & 80        & 5        \\
$\NV$     & 70        & 3        \\
$\OVI$    & 60        & 1        \\
$\MgII$   & 95        & 6        \\
\hline
\end{tabular}
\caption{The calculated lags (CCF peaks) of Model C for the smoothed
continuum light curve of PG0844, shown in Fig.~\protect\ref{f:PG}. The CCF was
calculated without the first 100 days of the line light curve. The CCF is
calculated with a 5 day resolution. The rank order of the lags (1 is the
shortest) is also displayed for convenience.}
\label{t:lags}
\end{table}

%%%%%%%%%%%%%%%%%%%%%%%%%%%%%%%%%%%%%%%%%%%%%%%%%%%%%%%%%%%%%%%%%%%%%%%%
%%%%%%%%%%%%%%%%%%%%%%%%%%%%%%%%%%%%%%%%%%%%%%%%%%%%%%%%%%%%%%%%%%%%%%%%

\section{Discussion}
\label{s:discuss}

   In papers I and II, it was shown that the BS model can successfully
account for two main features of AGN spectra: the emission line ratios and
their profiles. Here, I study the line reverberation of the BS model and
check whether it can also account for this aspect of AGN properties. I
focus on the two trends that emerge from the available observational
results: the positive correlation between the AGN luminosity and the
typical time lag scale and the anti-correlation between the line time lag
and its degree of ionization.
   
   Several difficulties stand in the way of a direct comparison of the BS
model predictions with observed line reverberation results. From the
observational side, the picture is still unclear since the available sample
of adequately sampled AGN is small and spans only a limited range in
luminosity. A reliable statistical analysis of the time lags is notoriously
difficult, and this is compounded by the fact that the quoted time lags in
the various campaigns were derived by a variety of statistical procedures,
each with its own biases. The strong dependence of the time-lag on the
characteristic continuum variability is another factor that has yet to be
taken into account. 

   From the theoretical side, there is the difficulty that there is some
degeneracy among the BS model parameters that makes it possible to have
similar emissivity maps for different luminosities (see
section~\ref{s:model}), and therefore the luminosity of this AGN model is
not well defined. This was not a serious problem for comparing the BS model
line ratios and profiles with the the observed ones, because
observationally, these do not appear to be strongly correlated with the AGN
luminosity \cite{BNW}.  This is not the case with the time lag results, as
there is an indication of a correlation between the time lag and luminosity
\cite{Maoz3,Kaspi2}.

   To date, published time lag results, which are based on the Balmer
lines, exist only for 9 AGN (NGC 5548 \protect\cite{Korista}; NGC 4151
\protect\cite{Kaspi,Maoz2}; NGC 3227 \protect\cite{Salamanca}; NGC 3783
\protect\cite{Stirpe}; AKN 120 \protect\cite{Peterson,PG}; MRK 279
\protect\cite{Maoz,Stirpe2}; NGC 3516 \protect\cite{Wanders}; MRK 590
\protect\cite{Peterson3}; NGC 4593 \protect\cite{Dietrich2}). These AGN
have estimated luminosities\footnote{The luminosity is integrated in the
range 0.1--1$\mu$, based on the measured flux at $\sim5000$ \AA\/ and the
assumption of a power-law continuum $F_\nu\propto\nu^{-1/2}$, $H_0 = 75$ km
s$^{-1}$ Mpc$^{-1} $ and $q_0 = 1/2$.} in the range $6.5\e{42}$ to
$1.8\e{44}$ erg s$^{-1}$ and $\Hb$ time lags in the range 4 to 28 d. This
sample suggests a weak positive correlation between the luminosity and the
$\Hb$ time lag. A major improvement in testing this correlation is offered
by new results from a line reverberation campaign on a sample of QSO with
luminosities from $4\e{44}$ to $3\e{46}$ erg s$^{-1}$.  Preliminary
estimates of the time lags confirm that the correlation is indeed real and
yield $\Hb$ time lags of approxiamtely 100 d for the luminosity range 2 to
$5\e{45}$ erg s$^{-1}$ \cite{Kaspi2}.

   Because it is possible that some of the correlation between the time lag
and luminosity is caused by a correlation between the continuum variability
time scales and the luminosity, I use in the simulations the optical
continuum of PG0844 (one of the QSO in this new sample), which is close in
its luminosity to that of model C. In doing so, I am assuming that the
ionizing light curve is similar to the optical one. The luminosity of
PG0844, when estimated by the same procedure as above, is $4.7\e{44}$ erg
s$^{-1}$ and that of model C is $2.3\e{44}$ erg s$^{-1}$. I thus expect the
lag of model C to be intermediate between those of the observed Sy1 sample
(lags $\la 30$ d for $L \la 2\e{44}$ erg s$^{-1}$) and those of the new
QSO sample (lags $\sim 100$ d for $L \sim 4\e{45}$ erg s$^{-1}$). The model
$\Hb$ and $\Ha$ lags of 60 and 75 d, respectively, (Table~\ref{t:lags})
indeed place it in this intermediate range (as do the time lags of all the
other lines). Thus, the time lag of BS model C is consistent with these
observations.

   The BS model time lags display to some extent the trend for an inverse
correlation between the lag and the degree of ionization (`time lag line
segregation'), which is observed in a few low-luminosity Sy1s.  Two groups
of lines can be identified among the 11 broad emission lines which were
calculated. The relatively high ionization species, $\HeII$, $\NV$ and
$\OVI$, have narrower transfer functions and shorter lags than the
relatively low ionization species, $\CIII$, $\CIIIb$ and $\MgII$
(Fig.~\ref{f:psi} and table~\ref{t:lags}). There are two notable
differences between the model results and the observations.  The $\CIV$ has
a large time lag, which is comparable to those of the low ionization lines
and the $\Hb$ has a short time lag, comparable to that of the high
ionization lines.

   There are reasons to suspect that line reverberation time lags of
low-luminosity AGN can not be simply extrapolated to higher luminosities.
The results presented in Fig.~\ref{f:gcomp} suggest that any detailed
ordering of the line time lags, especially those with short time lags,
strongly depends on the typical variability time scales of the ionizing
continuum. This may explain the seeming $\CIV$ discrepancy. The $\Hb$
discrepancy may also be related to the ``$\La/\Hb$ problem''. All
photoionization models of the BLR consistently under-predict the Balmer
lines emission. This is one of the outstanding problems in AGN research,
which remains unsolved despite extensive efforts. The most likely
explanation is a combination of reddening and inaccurate treatment of the
line transfer, which arises from the local nature of the escape probability
method. This simplified approach may be inadequate for the extreme $\La$
and $\Ha$ optical depths that are typical of the BLR. Detailed line
transfer calculations, which deal separately with the different frequency
bins in the line profiles, are needed (see Netzer 1990 for a review). Such
calculations are beyond the scope of this paper. This situation casts
doubt on the reliability of the calculated model $\Hb$ time lags. Note,
however, that the agreement between the model time lag scale and the
observed time lag--luminosity correlation does not hang only on the $\Ha$
and $\Hb$ time lags but is also supported by the time lags of the other
lines.

   The time lag line segregation of the BS model is not as pronounced as that
found in the NGC 5548 campaign \cite{Clavel,Peterson2,Dietrich}. A similar
trend is also seen with respect to the line profiles. The observed profiles
display line to line differences (`profile line segregation'), which is
more pronounced than that of the model. This may indicate that the BS
fraction is even more centrally concentrated than the assumed $\fBS \propto
r^{-2}$.

   This rough agreement of the model with the observations reinforces two
of the conclusions that followed from the line profile study in paper II.
The line segregation in the profile widths and time lags is yet another
indication that the BSs can not be Comptonized (paper I). Comptonization
implies that the BS wind boundary is determined by the ionizing flux and is
at a constant ionization parameter, irrespective of the distance from the
black hole. This leads to similar emissivity maps for all lines and hence
to similar profiles and lags, contrary to what is observed. The line
profiles and lags also require that the fraction of BSs in the stellar
population, $\fBS$, fall off at least as $r^{-2}$. A less concentrated BS
distribution results in significantly larger lags. For example, BS model B
of paper II, which is identical to model C apart for having $\fBS = {\rm
const.}$, yields a $\Hb$ lag of 210 d for the PG0844 continuum. This is in
clear contradiction with the observations.

   In this paper I studied one particular BS model (model C), which is
specified by a particular choice of the AGN dynamical age, the BS envelope
structure and the ionizing spectrum. Other choices of these parameters will
result in other model predictions. This work is part of a series of
feasibility studies aimed at checking whether one BS model can
simultaneously reproduce the main broad line properties of a typical
AGN. Model C was shown to reproduce the line ratios and line profile, and
was therefore also in the focus of this line reverberation study. Direct
comparison of model C with observed line reverberation time lags must await
future monitoring campaigns of bright Sy1s. The question whether a BS model
of a low-luminosity Sy1 is consistent with the detailed line reverberation
data available on such AGN is part of the larger question of whether the BS
model can be extended to the full range of observed AGN luminosities. This
will be explored in a future paper. In the meanwhile, I conclude that BS
models, similar to those discussed here, are consistent with current line
reverberation data on AGN in the bright Sy1 / low-luminosity QSO range.

\hfill\\
{\bf Acknowledgments}\hfill\\

   I am very grateful to Hagai Netzer for his advice and careful reading of
the draft. Shai Kaspi's help and permission to use unpublished results are
much appreciated. This research was partially supported by the Israel
Science Foundation administered by the Israel Academy of Sciences and
Humanities and by the Jack Adler chair of Extragalactic Astronomy at Tel
Aviv University.

\end{document}